\documentclass[conference]{IEEEtran}
\IEEEoverridecommandlockouts
\usepackage{cite}
\usepackage{graphicx}
\usepackage{textcomp}
\usepackage{xcolor}

\usepackage{booktabs} 
\usepackage[utf8]{inputenc}
\usepackage{multirow}
\usepackage{hyperref}
\usepackage{amsmath,amssymb,amsthm}
\usepackage{amsfonts}
\usepackage{xspace}
\usepackage{float}
\usepackage{subcaption}
\usepackage{caption}
\usepackage{bm}
\usepackage[ruled, vlined, linesnumbered, nofillcomment]{algorithm2e}
\usepackage{mathtools}
\usepackage{stmaryrd}

\newcommand{\spara}[1]{\smallskip\noindent{\bf{#1}}}

\newcommand{\bfu}{\ensuremath{\mathbf{u}}\xspace}

\newtheorem{problem}{Problem}
\newtheorem{theorem}{Theorem}
\newtheorem{lemma}[theorem]{Lemma}
\newtheorem{proposition}{Proposition}
\theoremstyle{definition}
\newtheorem{definition}{Definition}[section]

\newcommand{\set}[1]{\left\{#1\right\}}
\newcommand{\pr}[1]{\left(#1\right)}

\newcommand{\spr}[1]{\left[#1\right]}

\newcommand{\vect}[1]{\spr{#1}}

\newcommand{\reals}{{\mathbb{R}}}
\newcommand{\positivereals}{{\reals_{+}}}

\newcommand{\expectation}{\ensuremath{\mathbb{E}}\xspace}

\newcommand{\Gaussian}{\ensuremath{\mathcal{N}}\xspace}
\newcommand{\bigO}{\ensuremath{\mathcal{O}}\xspace}
\newcommand{\np}{\ensuremath{\mathbf{NP}}\xspace}
\newcommand{\poly}{\ensuremath{\mathbf{P}}\xspace}

\newcommand{\yes}{\texttt{yes}\xspace}
\newcommand{\no}{\texttt{no}\xspace}

\newcommand{\bfz}{\ensuremath{\mathbf{z}}\xspace}
\newcommand{\bfs}{\ensuremath{\mathbf{s}}\xspace}

\newcommand{\bfx}{\ensuremath{\mathbf{x}}\xspace}
\newcommand{\bfxopt}{\ensuremath{\mathbf{x}^*}\xspace}
\newcommand{\bfxbin}{\ensuremath{\bar{\mathbf{x}}}\xspace}
\newcommand{\xbinelement}{\ensuremath{\bar{{x}}}\xspace}
\newcommand{\bfxbest}{\ensuremath{\mathbf{x}_\mathit{b}}\xspace}
\newcommand{\bfy}{\ensuremath{\mathbf{y}}\xspace}
\newcommand{\bfr}{\ensuremath{\mathbf{r}}\xspace}
\newcommand{\bfq}{\ensuremath{\mathbf{q}}\xspace}
\newcommand{\bfone}{\ensuremath{\mathbf{1}}\xspace}
\newcommand{\bfzero}{\ensuremath{\mathbf{0}}\xspace}

\newcommand{\selement}{\ensuremath{s}\xspace}
\newcommand{\xelement}{\ensuremath{x}\xspace}

\newcommand{\laplacian}{\ensuremath{\mathbf{L}}\xspace}
\newcommand{\Lelement}{\ensuremath{L}\xspace}

\newcommand{\matrixQ}{\ensuremath{\mathbf{Q}}\xspace}

\newcommand{\matrixX}{\ensuremath{\mathbf{X}}\xspace}
\newcommand{\matrixXopt}{\ensuremath{\mathbf{X}^*}\xspace}
\newcommand{\matrixV}{\ensuremath{\mathbf{V}}\xspace}
\newcommand{\matrixI}{\ensuremath{\mathbf{I}}\xspace}
\newcommand{\matrixM}{\ensuremath{\mathbf{M}}\xspace}
\newcommand{\Melement}{\ensuremath{M}\xspace}
\newcommand{\matrixY}{\ensuremath{\mathbf{Y}}\xspace}
\newcommand{\matrixP}{\ensuremath{\mathbf{P}}\xspace}
\newcommand{\Pelement}{\ensuremath{P}\xspace}
\newcommand{\matrixD}{\ensuremath{\mathbf{D}}\xspace}
\newcommand{\Delement}{\ensuremath{D}\xspace}
\newcommand{\matrixA}{\ensuremath{\mathbf{A}}\xspace}
\newcommand{\Aelement}{\ensuremath{A}\xspace}

\newcommand{\graph}{\ensuremath{G}\xspace}
\newcommand{\vertices}{\ensuremath{V}\xspace}
\newcommand{\novertices}{\ensuremath{n}\xspace}
\newcommand{\edges}{\ensuremath{E}\xspace}
\newcommand{\noedges}{\ensuremath{m}\xspace}
\newcommand{\eweight}{\ensuremath{w}\xspace}
\newcommand{\vweight}{\ensuremath{\mathbf{b}}\xspace}
\newcommand{\vwelement}{\ensuremath{b}\xspace}
\newcommand{\diversity}{\ensuremath{\eta}\xspace}
\newcommand{\Diag}{\ensuremath{\mathit{Diag}}\xspace}
\newcommand{\diag}{\ensuremath{\mathit{diag}}\xspace}
\newcommand{\trace}{\ensuremath{\mathit{Tr}}\xspace}
\newcommand{\opt}{\ensuremath{\mathit{Opt}}\xspace}
\newcommand{\rank}{\ensuremath{\mathit{rank}}\xspace}
\newcommand{\card}{\ensuremath{\mathit{card}}\xspace}

\newcommand{\gbound}{\ensuremath{R}\xspace} 
\newcommand{\lambdamax}{\ensuremath{{\lambda_{\max}}}\xspace}
\newcommand{\sparsity}{\ensuremath{r}\xspace}
\newcommand{\iter}{\ensuremath{I}\xspace}
\newcommand{\budget}{\ensuremath{k}\xspace}
\newcommand{\maxcut}{{\sc\large max-cut}\xspace}
\newcommand{\subsetsum}{{\sc\large sub\-set-sum}\xspace}
\newcommand{\qbk}{{\sc\large qbk}\xspace}

\newcommand{\sdp}{{\sc\large sdp}\xspace}
\newcommand{\sdpqbk}{{\sc\large sdp-qbk}\xspace}
\newcommand{\bsdp}{{\sc\large p}\xspace}
\newcommand{\sdpr}{{\sc\large r}\xspace}
\newcommand{\glover}{{\sc\large Glover}\xspace}
\newcommand{\qkp}{{\sc\large qkp}\xspace}

\newcommand{\sdpalgo}{{\tt SDP-Relax}\xspace}
\newcommand{\exactalgo}{{\tt IQP}\xspace}
\newcommand{\heuristicalgo}{{\tt S-Greedy}\xspace}
\newcommand{\greedyalgo}{{\tt I-Greedy}\xspace}
\newcommand{\gloveralgo}{{\tt Glover}\xspace}

\newcommand{\karate}{{\sf\small Karate}\xspace}
\newcommand{\karated}{{\sf\small Karate-D}\xspace}
\newcommand{\books}{{\sf\small Books}\xspace}
\newcommand{\booksd}{{\sf\small Books-D}\xspace}
\newcommand{\twitter}{{\sf\small Twitter100}\xspace}
\newcommand{\twitterbig}{{\sf\small Twitter}\xspace}
\newcommand{\twitterd}{{\sf\small Twitter100-D}\xspace}
\newcommand{\blogs}{{\sf\small Blogs}\xspace}
\newcommand{\elections}{{\sf\small Elections}\xspace}
\def\BibTeX{{\rm B\kern-.05em{\sc i\kern-.025em b}\kern-.08em
    T\kern-.1667em\lower.7ex\hbox{E}\kern-.125emX}}
\begin{document}

\title{Tell me something my friends do not know: \\
Diversity maximization in social networks}

\author{
\IEEEauthorblockN{Antonis Matakos}
\IEEEauthorblockA{\textit{Department of Computer Science} \\
\textit{Aalto University}\\
Espoo, Finland \\
firstname.lastname@aalto.fi}
\and
\IEEEauthorblockN{Aristides Gionis}
\IEEEauthorblockA{\textit{Department of Computer Science} \\
\textit{Aalto University}\\
Espoo, Finland \\
firstname.lastname@aalto.fi}
}

\maketitle

\begin{abstract}
Social media have a great potential to improve information dissemination in our society, 
yet, they have been held accountable for a number of undesirable effects,
such as polarization and filter bubbles.
It is thus important to understand these negative phenomena and 
develop methods to combat them.
In this paper we propose a novel approach
to address the problem of breaking filter bubbles in social media.
We do so by aiming to maximize the diversity of the information exposed to 
connected social-media users.
We formulate the problem of maximizing the diversity of exposure
as a quadratic-knapsack problem.
We show that the proposed diversity-maximization problem is 
inapproximable, and thus, we resort to polynomial non-approximable algorithms,  
inspired by solutions developed for the quadratic-knapsack problem, 
as well as scalable greedy heuristics.
We complement our algorithms with instance-specific upper bounds, 
which are used to provide 
empirical approximation guarantees for the given problem instances. 
Our experimental evaluation shows that a proposed greedy algorithm followed by randomized local search
is the algorithm of choice given its quality-vs.-efficiency trade-off. 
\end{abstract}


\section{Introduction}

Social media play a critical role in today's information society,
not only by connecting people with their friends,
but also as a means of news dissemination.
A recent survey estimates that 6 out of 10 adults in the US
get their news on social media~\cite{pew}.
Although initially it appeared that
social media can contribute to the democratization of content generation and distribution,
most recently, a series of negative effects and undesirable phenomena have emerged,
such as \emph{filter bubbles}, \emph{polarization}, \emph{fake news}, and more.
An indication for the extent of the problem
can be seen by Facebook's own admission that
\emph{social media can have the unintended consequence of corroding democracy}.%
\footnote{\url{http://money.cnn.com/2018/01/22/technology/facebook-democracy-social-media/index.html}, January 22, 2018}

Given these negative effects, a recent body of research has focused on solving different aspects of the problem.
Proposed approaches
include detecting ~\cite{akoglu14,garrett09echo,conover11political,plos1,garimella16quantifying, guerra13measure} and
reducing polarization of opinions~\cite{Matakos17, Musco17},
recommending users-to-follow and
content to bridge opposing views~\cite{Liao2014, garimella17reducing, munson10presenting},
and balancing information exposure in a social network~\cite{garimella17balancing}.

One particular aspect of the problem is the emergence of
\emph{filter bubbles} in social media~\cite{pariser11filter},
where individuals are connected to like-minded net-citizens or information sources of similar disposition,
becoming unexposed to information they disagree with,
and effectively isolated in their own cultural or ideological bubbles.

In this paper we propose a novel approach
to address the problem of breaking filter bubbles.
Our underlying assumption is that a filter bubble
is the \emph{lack of diversity in the information exposure}
among connected individuals in the social network.
We further assume that diversity of information exposure in the social network
can be increased by means of content recommendation.
What this means is that in addition to the
content that circulates in the network ``organically''
via shares and re-posts among users,
the social-media platform may consider to
strategically recommend suitably-chosen content to
selected individuals in order to increase diversity,
and thus, help bursting filter bubbles.%
\footnote{To increase transparency,
the social-media platform may need to distinctly mark,
or display in a specific format,
such recommendations so that users are aware of its presence and objective.
Additionally, users may need to opt-in receiving such recommendations.
We consider, however, that such design
considerations are orthogonal to our study
and beyond the scope of this paper.}

A desirable property for such a mechanism of diversity-enabling recommendations
is that the social-media platform should make a \emph{small amount of recommendations},
as these can be perceived as interventions to the organic operation of the network.
Thus, a natural question to ask is:
\emph{``given a fixed amount of content-recommendation activity,
which users should we target and what recommendations to make,
so as to maximally increase diversity?''}
Intuitively, we would like to target users who are part of filter bubbles
(their friends and themselves are exposed to similar content),
are influential among their connections,
and are likely to take into account the recommended content.

Our problem formulation puts together all these components.
Naturally, we need to make a number of modeling assumptions.
We assume that we know the structure of the social graph,
and the influence that users exert on each other,
expressed as edge weights on the social graph.
We also assume that we can quantify the information exposure
of each individual as a number in a pre-specified range
(we use [-1,1]),
e.g.,
expressing the grade of the content consumed by an
individual in a continuous spectrum, for example, \emph{conservative--liberal}.
Finally, we assume that we can estimate the degree to which
an individual is likely to take into account a specific recommendation,
e.g.,
the probability to re-post the recommended item.
Although estimating these parameters is orthogonal to our study,
it is easy to see that one can develop simple proxies for them, using
data available in the social-media platform.

\begin{figure}[t]
\centering
\begin{subfigure}[b]{0.23\textwidth}
\includegraphics[width=\textwidth]{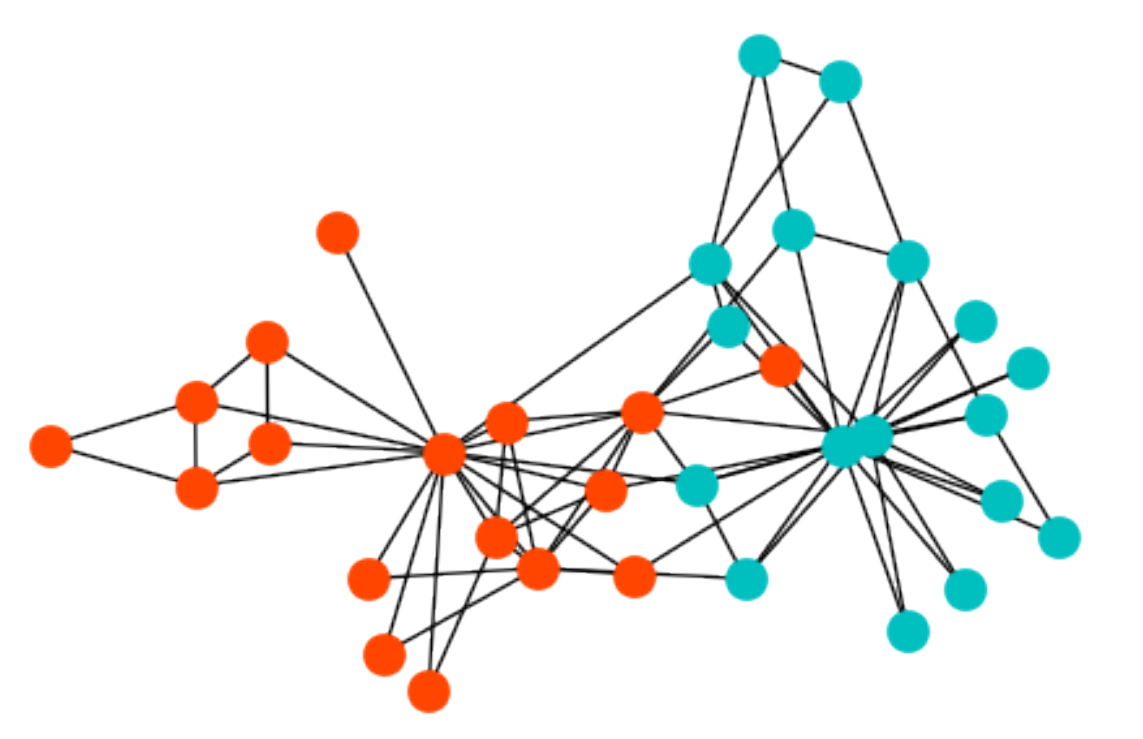}
\caption{\label{fig:toy-example-a}Echo-chamber graph} 
\end{subfigure}
\begin{subfigure}[b]{0.23\textwidth}
\includegraphics[width=\textwidth]{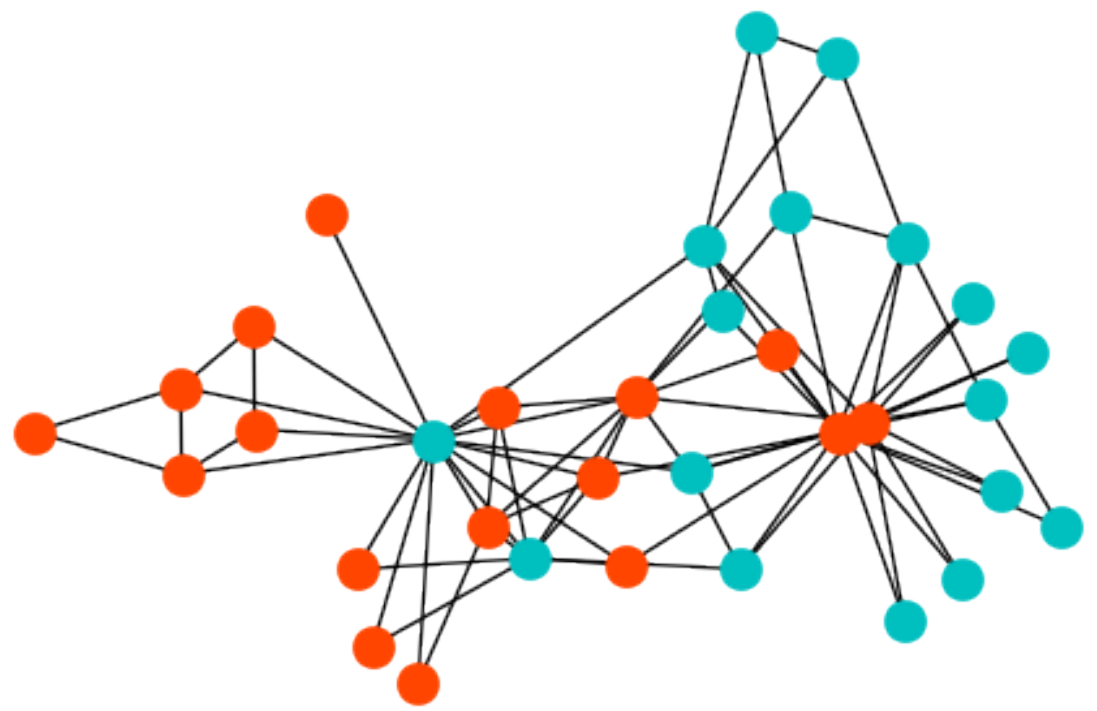}
\caption{\label{fig:toy-example-b}Graph with diversified exposure}
\end{subfigure}
\caption{\label{fig:toy-examples}
Toy graph with different exposure assignments}
\end{figure}

\smallskip
\noindent
{\bf Example:}
A toy example demonstrating our concept is shown in Figure~\ref{fig:toy-examples},
using the Karate-club network, which is known to contain two communities.
The colors on the nodes represent different exposure levels,
say two different ``news diets'' that the network users consume.
In Figure~\ref{fig:toy-example-a}
each community has different exposure level,
leading to a network with echo chambers and no diversity.
In Figure~\ref{fig:toy-example-b} we depict the optimal solution
to our problem,
where we ask for the best $\budget=4$ users,
to change their exposure and maximize the total network diversity---%
assuming that all users opt-in to receive alternative news diets and
the user cost is constant.
In this simple example the algorithm picks the two hubs of each community.

\smallskip
From the technical point of view,
we formulate the problem of maximizing diversity of exposure
as a special case of the quadratic-knapsack problem (\qkp)~\cite{gallo1980quadratic}.
Our first result shows that the diversity maximization problem is not only \np-hard,
but also \np-hard to approximate within a multiplicative factor.
Thus, we study a number of polynomial algorithms
inspired by the quadratic-knapsack formulation,
such as methods based on semidefinite-programming (\sdp) relaxation and linearization techniques.
We also propose two scalable greedy algorithms,
which take advantage of the special structure of our problem.

Our results show that the \sdp-based algorithm is the best performing
on a diverse range of settings,
followed very closely by one of the greedy methods.
This is very useful because while the \sdp algorithm is expensive,
the greedy has linear complexity with respect to the number of nodes in the network,
and thus, has excellent scalability properties.

Our relaxation provides upper bounds on the quality of solution.
In addition we propose alternative upper bounds
with varying trade-offs of tightness-vs.-efficiency.
All these bounds allow us to obtain empirical approximation guarantees
for given problem instances.
For instance, for the problem instances used in our experiments,
we are able to assert that our algorithms give solutions
with typical approximation factor between 1.5 and 2.5;
despite the problem being \np-hard to approximate.

In summary, in this paper we make the following contributions:
\begin{itemize}
\item Inspired by the problem of breaking filter bubbles,
we formulate the problem of maximizing the diversity of exposure,
as a special case of the quadratic-knapsack problem.
\item
We prove that the diversity maximization problem is
\np-hard to approximate within a multiplicative factor.
\item
We study several algorithms for the problem,
including an \sdp-based algorithm,
an algorithm based on linearization,
and two greedy methods.
\item
We develop upper bounds with different trade-offs of tightness-vs.-efficiency,
which provide empirical approximation guarantees for given problem instances.
\item
We present an extensive experimental evaluation
that provides evidence for the best-performing methods,
and quality-vs.-efficiency trade-offs.
\end{itemize}

The rest of the paper is organized as follows.
We start our presentation by reviewing the related work, in Section~\ref{section:related}.
We then present our notation in Section~\ref{section:diversity-index},
and we formally define the diversity maximization problem in Section~\ref{section:formulation}.
The \np-hardness proof is also presented in Section~\ref{section:formulation}.
In section~\ref{section:algorithms} we discuss algorithms for the binary version of the problem,
and we present upper bounds for the optimal solution.
The extension of the diversity-maximization problem to the continuous case
is discussed in Section~\ref{section:continuous}.
We present our experimental evaluation in Section~\ref{section:experiments},
and we conclude in Section~\ref{section:conclusions} by offering our final remarks
and suggestions for future work.

\section{Related work}
\label{section:related}

Our work relates to the emerging line of work on
breaking filter bubbles and reducing polarization on social media.
To the best of our knowledge, this is the first work
to approach this problem from the point of view of
increasing diversity of information exposure,
and formulating it as a quadratic knapsack-style problem.

\smallskip
\noindent
\textbf{Detecting polarization:}
Recently, a significant body of work has emerged that focuses on measures for
characterizing polarization in online social
media~\cite{Matakos17, akoglu14, conover11political, garimella16quantifying, guerra13measure}.
These works consider mainly the structure in social-media interactions
and quantify polarization or compute node polarity scores
using network-based techniques.
Other papers study the emergence of polarization on various opinion-formation models:
Dandekar et al.~\cite{goel} generalize DeGroot's model to account for \emph{biased assimilation},
while Vicario et al.~\cite{DBLP:journals/corr/VicarioSCSQ16}
propose a variant of the \emph{bounded-confidence model},
where discordant edges are rewired and two opposing opinion clusters emerge.

\smallskip
\noindent
\textbf{Reducing polarization:}
Given the negative effects of fragmentation,
there has been recent work that focuses on methods for
reducing polarization~\cite{Matakos17, Musco17,garimella17reducing}.
Matakos et al.~\cite{Matakos17} study the problem of convincing a set of individuals
to adopt a neutral opinion and act as mediators in the discussion.
Musco et al.~\cite{Musco17} study a similar problem,
albeit with the dual objective of minimizing both polarization and disagreement among individuals.
Garimella et al.~\cite{garimella17reducing}
consider the problem of introducing new edges
between the two sides of a controversy,
so as to reduce polarization.

There are two key differences between these works and our approach.
First, while these works focus on minimizing other measures of polarization,
our aim is to maximize the diversity of content that an individual is exposed.
Second, while these works consider how to affect user opinions, we only consider the \emph{exposure} of a user.
We consider our setting more realistic since in practice, it is difficult to know the opinions of the users.

A complementary line of work studies mechanisms that expose social-media users
to content that is not aligned
with  their prior beliefs~\cite{munson10presenting,vydiswaran15overcoming,Liao2014}.
While these works focus on \emph{how} to present information to users,
addressing issues of interface and incentives,
our work addresses the question of \emph{who} to approach with the new information.

\smallskip
\noindent
\textbf{Quadratic knapsack:}
Our formulation maps the diversity maximization problem
to a special case of the quadratic-knapsack problem (\qkp). 
The general form of \qkp was introduced by Gallo et al.~\cite{gallo1980quadratic}.
A classical technique to solve 0--1 quadratic problems is to
\emph{linearize} them by introducing auxiliary variables and transforming the problem to an
integer linear program ({\sc ilp}) formulation.
Glover et al.~\cite{Glover75} presented a concise way to rewrite a 0--1
quadratic problem as an equivalent 0--1 linear program
with only $n$ auxiliary variables and $4n$ constraints.

Of particular interest are \emph{semidefinite programming} (\sdp) techniques,
as they can yield tighter relaxations than using linearization,
albeit at the expense of higher running time.
Therefore, central to our work is the methodology of Helmberg et al.~\cite{Helmberg96asemidefinite},
who approach \qkp by introducing
a series of \sdp relaxations of increasing tightness.
Additionally, to strengthen the formulation a number of inequalities defining the \qkp polyhedron
(called the boolean quadric polytope) have been studied~\cite{Padberg89}.
On a high level, our problem is also related to the \maxcut problem,
which has been shown to admit an approximation ratio of 0.878~\cite{Goemans95},
using semidefinite programming.
However, as we prove shortly, our problem is harder as it admits no polynomial approximation guarantee.

\section{Diversity Index}
\label{section:diversity-index}

Consider a social network represented as a graph $\graph = (\vertices, \edges, \eweight)$,
where the node set \vertices represents a set of individuals,
the edge set \edges represents social connections,
and $\eweight:\edges\rightarrow\positivereals$ is a weight function
that represents the strength of the social connections. 
The weight of an edge $(i,j)\in\edges$ is denoted by $\eweight_{ij}$.
The number of nodes and edges 
are denoted by \novertices and \noedges, respectively.

We write \matrixA to denote the adjacency matrix of the graph \graph,
whose entry $\Aelement_{ij}$ is equal to $\eweight_{ij}$ if $(i,j)\in\edges$,
and 0 otherwise.
We assume that the graph \graph is undirected, and so \matrixA is symmetric.
We also denote by \matrixD the diagonal matrix
whose $i$-th diagonal entry is equal to the weighted degree of individual $i$,
i.e., $\Delement_{ii} = \sum_j \eweight_{ij}$.
The \emph{Laplacian matrix} of the graph \graph is
$\laplacian = \matrixD - \matrixA$.

We assume that each individual $i\in\vertices$ has an overall \emph{exposure} to content
represented by a value $\selement_i$.
The value $\selement_i$ may represent the leaning of the content individual $i$
is exposed to on an issue, as measured by endorsed or shared news articles.
Without loss of generality we assume that $\selement_i\in\spr{-1,1}$.
We also consider the discrete case where $\selement_i\in\set{-1,1}$.
The vector of exposure for all the individuals in the graph
is denoted by 
$\bfs = \vect{\selement_i}_{i\in\vertices}$.

Given a vector of exposure \bfs for the individuals of a network
we are interested in measuring the \emph{network diversity}
with respect to~\bfs.
Intuitively,
high diversity should indicate that many individuals tend to have
different exposure than that of their social connections.
Furthermore, diversity should account for the strength of social connections.
These considerations motivate our definition of
the \emph{diversity index} in a social network.

\begin{definition}[Diversity index]
Given a graph $\graph = (\vertices, \edges, \eweight)$,
and a vector of exposure
$\bfs \in \spr{-1, 1}^{\novertices}$ for the individuals in \vertices,
the \emph{diversity index} $\diversity(\graph, \bfs)$
of the graph \graph with respect to the exposure vector \bfs
is defined as :
\begin{equation}
\label{eq:index}
\diversity(\graph, \bfs) = \sum_{(i,j)\in \edges} \eweight_{ij}\pr{\selement_i-\selement_j}^2.
\end{equation}
\end{definition}
An equivalent way of writing Equation~(\ref{eq:index}),
using the laplacian \laplacian of the graph \graph is:
\begin{equation}
\label{eq:computation}
\diversity(\graph, \bfs)
	= \sum_{(i,j)\in \edges} \eweight_{ij}\pr{\selement_i-\selement_j}^2
	= \bfs^\top\matrixD\,\bfs - \bfs^\top \matrixA\, \bfs
	= \bfs^\top\laplacian\,\bfs.
\end{equation}
Higher value for the diversity index indicates more diverse networks.

In the rest of a paper,
given a vector \bfx we write $\Diag(\bfx)$
to denote the \emph{matrix} with \bfx as its diagonal, and
given a matrix \matrixX we write $\diag(\matrixX)$
to denote the \emph{vector} corresponding to the diagonal of \matrixX.
As it is common, we denote by $\trace(\matrixX)$ the \emph{trace} of a matrix \matrixX.
Finally, we write $\matrixX \succeq 0$ to denote that \matrixX is a positive semidefinite matrix.

\section{Problem formulation}
\label{section:formulation}

In this paper our main focus is on the discrete case where exposure $\selement_i$ is either
$-1$ or $1$. This corresponds to the case where discussions are characterized by two dominant and opposing perspectives, which, exacerbated by filter bubbles often leads to polarization and lack of diversity of exposure: some examples are the fragmentation into liberals vs conservatives, brexit vs bremain, right-wing vs left-wing. Additionally, this assumption leads to a more attractive formulation, while being at least as challenging computationally. We also investigate the continuous case but provide it more as an extension.

Our goal is to maximize the \textit{diversity index}
$\diversity(\graph,\bfs)$ of a graph \graph,
assuming that we know the current exposure vector \bfs.
We consider maximizing the diversity index by
selecting individuals and ``flipping'' the leaning of their exposure
(from $-1$ to $1$, or vice versa), under a budget constraint.

As mentioned in the introduction,
changing the exposure of a individual corresponds to
recommendations that an individual possibly opts-in.
The issue of interface and communication with the individuals
is of independent interest and we consider it orthogonal to our work.

Given an exposure vector \bfs,
after changing the exposure of \budget individuals,
the new exposure vector can be written as $\bfy=\bfs-\bfs'$,
where $\bfs'$ is a sparse vector with at most \budget non-zero elements.
In particular, $\selement'_i=-2$ if $\selement_i=-1$ and $\selement'_i=2$ if $\selement_i=1$.
Alternatively, we can write the new exposure vector as
$\bfy=\bfs-2\,\Diag(\bfs)\,\bfx$,
where $\bfx \in \{0,1\}^\novertices$ is an indicator vector
with $\xelement_i=1$ if the exposure of the $i$-th individual has changed and
$\xelement_i=0$ otherwise.
This formulation highlights the nature of the problem as a variable-selection problem.

We consider a knapsack-type constraint for \bfx with a weight vector~\vweight,
where $\vwelement_i$ expresses the \emph{cost} in altering the exposure of individual $i$
(for example, some individuals may have a strong predisposition towards certain issues).
We define the following problem:

\begin{problem}
\label{problem:problem1}
Given a graph $\graph=(\vertices,\edges)$, an \textit{exposure} vector $\bfs$,
a node-weight vector \vweight,
a budget \budget,
we ask to find a binary vector \bfx,
such that the knapsack constraint $\vweight^\top\bfx\leq k$ is satisfied
and the resulting diversity index $\eta(G,\bfy)$ is maximized,
where $\bfy=\bfs-2\,\Diag(s)\,\bfx$.
\end{problem}

We now formulate the corresponding optimization problem.
Using the definition of the diversity index from Equation~(\ref{eq:computation}) we have
\begin{eqnarray*}
\label{eq:AP}
\bfy^\top\laplacian\,\bfy
& = & \left(\bfs-2\,\Diag(\bfs)\,\bfx\right)^\top\laplacian \left(\bfs-2\,\Diag(\bfs)\,\bfx\right) \\
& = & \bfs^\top\laplacian\,\bfs+4\,\bfx^\top\,\Diag(\bfs)\,\laplacian\,\Diag(\bfs)\,\bfx \\
&   &  - 4\,\bfs^\top\laplacian\,\Diag(\bfs)\,\bfx.
\end{eqnarray*}

Recall that we want to maximize this quantity.
The term $\bfs^\top\laplacian\,\bfs$ is a constant and has no influence on the maximization,
therefore can be removed from the objective.
For convenience, we define
\[
\matrixQ=\Diag(\bfs)\,\laplacian\,\Diag(\bfs), \;
\bfq=\bfs^\top\laplacian\,\Diag(\bfs),
\mbox{ and }
\]
\begin{equation}
\label{equation:P}
\matrixP=\matrixQ-\Diag(\bfq).
\end{equation}

Note that \matrixP is constant and does not depend on the optimization variable~\bfx.
We take advantage of the fact that $\bfx =\bfx^2$
to write our program in purely quadratic form.
We define the following quadratic binary knapsack (\qbk) problem:
\begin{equation}
\tag{\qbk}\label{problem:QBK}
\begin{aligned}
  &\max & & \bfx^\top\matrixP\,\bfx \\
  & \text{subject to} & &  \vweight^\top\bfx\leq k,  \quad \vweight \in \reals^\novertices \\
  & & &  \bfx \in \set{0,1}^\novertices.
\end{aligned}
\end{equation}

 It is not difficult to see that \qbk is \np-hard,
 by a simple transformation from {\maxcut}.
 It suffices to set $\vwelement_i=0$, and the problem reduces to:
\begin{equation*}
\begin{aligned}
     &\max & & \bfy^\top\laplacian\,\bfy \\
     & \text{subject to} & &  \bfy \in \set{-1,1}^\novertices, \\
\end{aligned}
\end{equation*}
which is equivalent to a general instance of {\maxcut}.
Despite this similarity, our problem is harder than {\maxcut}.
This is because {\maxcut} admits an approximation algorithm
with ratio 0.878~\cite{Goemans95},
while it can be shown that \qbk is \emph{inapproximable}.

\begin{proposition}
\label{proposition:inapproximable}
The \qbk problem is \np-hard to approximate.
\end{proposition}

The Proposition asserts that there cannot be a polynomial-time approximation algorithm
for the \qbk problem with a multiplicative approximation guarantee.
It does not preclude though the existence of an algorithm
with additive approximation guarantee.

\begin{proof}
We prove the Proposition 
by a reduction from {\subsetsum},
a known \np-complete problem~\cite{GJ79}.
An instance of {\subsetsum} consists of $n+1$ positive integers $m_1,...,m_n$ and~$M$.
The problem asks if there is a subset $S \subseteq \set{1,...,n}$ such that $\sum_{i \in S}m_i=M$.

Define $A = \sum_{i=1}^\novertices m_i$.

Given an instance of {\subsetsum},
we construct an instance of \qbk as follows:
the underlying graph \graph is a star graph with $n+1$ leaves.
The central node is node 0.
We assign weights to edges $(0,i)$ by
$\eweight_{0,i}=-m_i$, for $i=1,...,n$, and
weight $\eweight_{0,n+1}=M-A-1$ to edge $(0,n+1)$.
We set the budget $\budget=M$ and the node weights
$\vwelement_0=0$,
$\vwelement_i=m_i$, for $i=1,...,n$, and $\vwelement_{n+1}=M+1$ ---
for $\vwelement_{n+1}$ any other number bigger than $M$ also works,
since the goal is to make selection of this coefficient infeasible.
Finally we set $\selement_i=1$, for all $i=0,1,\ldots,n,n+1$,
so that the matrix $\matrixP$ is the laplacian $\laplacian$.
Observe that the resulting laplacian has the form
\[
\laplacian =
\left(
\begin{array}{ccccc}
-M+1       &  m_1   & \cdots  & m_n    & M-A-1 \\
m_1        & -m_1   & \cdots  &  0     & 0      \\
\vdots     & \vdots & \ddots  & \vdots & \vdots \\
m_n        &   0    & \cdots  & -m_n   & 0      \\
M-A-1 &   0    & \cdots  &   0    & -M+A+1 \\
\end{array}
\right)
.
\]

Now consider a binary vector $\bfx =\vect{\xelement_i}_{i=0,\ldots,n+1}$.
We interpret \bfx as a solution to \qbk, and
the coordinates $\vect{\xelement_1,\ldots,\xelement_n}$ as indicator variables
for a solution to \subsetsum.

First note that
due to the knapsack constraint $\vweight^\top\bfx\le M$,
since $\vwelement_{n+1}=M+1$ and since $\vwelement_{i}\ge 0$ for all $i$,
it is $\xelement_{n+1}=0$.

If $\xelement_{0}=0$,
then any feasible solution to \qbk can be at most 0,
and in fact the value 0 can be obtained by the feasible vector $\bfx=\bfzero$.

If $\xelement_{0}=1$,
let $S \subseteq \set{1,...,n}$ be the set of all other non-zero coordinates
in a feasible solution to \qbk.
The value of the solution is
$f = (-M+1) + 2\sum_{i\in S} m_i - \sum_{i\in S}m_i = \sum_{i\in S}m_i - M +1$.
Due to the knapsack constraint $\sum_{i\in S}m_i \le M$,
it follows that
if the answer to \subsetsum is \no we have $f\le 0$,
while
if the answer to \subsetsum is \yes then $f$ can obtain the value 1.

We conclude that the optimal value to \qbk is 1 if and only if the answer to \subsetsum is \yes,
while
the optimal value to \qbk is 0 if and only if the answer to \subsetsum is \no.

Furthermore,
any polynomial-time approximation algorithm with a finite (multiplicative)
approximation guarantee that could be used to solve \qbk,
will need to provide a non-zero value for \qbk
if any only if the answer to {\subsetsum} is \yes.
Thus, no such algorithm can exist, unless $\poly=\np$.
\end{proof}

\section{Algorithms}
\label{section:algorithms}

In this section we discuss the proposed algorithms
for the diversity-maximization problem,
and present upper bounds on the optimal solution.

We start by the observation that \qbk is a non-convex optimization problem
(due to \matrixP being not positive-semidefinite and the constraint $\bfx \in \set{0,1}^\novertices$),
which is expected
since convex problems can be solved in polynomial time.
We can show however, that we can still produce a convex semidefinite relaxation to this problem.
Such a relaxation forms the basis for our first algorithm.

\subsection{Semidefinite-programming relaxation}

Semidefinite-programming (\sdp) relaxations have long been studied in the optimization community.
The idea was introduced by Lovasz \cite{Lovasz79},
but it was arguably the seminal work of Goemans and Williamson for the \maxcut problem~\cite{Goemans95}
that brought {\sdp} relaxations into the spotlight.

We have noted that our problem (\qbk) is a non-convex 0--1 quadratic program
with a linear constraint on \bfx.
The general \textit{quadratic knapsack} problem,
first introduced by Gallo et al.~\cite{gallo1980quadratic},
is defined for an arbitrary symmetric matrix.
Since its introduction, a multitude of methods have been developed to solve the problem.
Of particular interest to us is the work of Helmberg et al.~\cite{Helmberg96asemidefinite},
in which they introduce a series of \sdp relaxations.

Now we present the \sdp relaxation to the \qbk problem.

\spara{Lift to a matrix variable.}
As is the common practice with semidefinite relaxations,
we ``lift'' the program to the space of square matrices.
In particular, we lift vector \bfx to a matrix \matrixX
by introducing the constraint $\matrixX=\bfx\,\bfx^\top$.
This constraint is equivalent to $\matrixX$ having rank~1 and being positive semidefinite.
However, the rank-1 constraint is not convex,
thus, we relax it to $\matrixX\succeq \bfx\,\bfx^\top$.
From the Schur complement this is equivalent to
\[
\begin{bmatrix}
\matrixX & \bfx \\
\bfx^\top & 1
\end{bmatrix}
\succeq 0.
\]
Observe that the constraint $\matrixX - \bfx\,\bfx^\top \succeq 0$ implies also that $\matrixX \succeq 0$.

\spara{Objective function.}
The objective of the \sdp relaxation is written as a function of the new variable \matrixX
in the trace form $\trace(\matrixP\,\matrixX)$,
since
$\bfx^\top\matrixP\,\bfx=\trace(\bfx^\top\matrixP\,\bfx)=\trace(\matrixP\,\bfx\,\bfx^\top)$.

\spara{Integrality constraint.}
The integrality constraint $\bfx\in\set{0,1}^\novertices$ can be written as $\bfx^2=\bfx$.
In the \sdp relaxation we write this as $\diag(\matrixX)=\bfx$.
We note that the polytope corresponding to the 0--1 quadratic optimization problem
is called the boolean quadric polytope~\cite{Padberg89}.
The boolean quadric polytope has a number of facet-defining inequalities,
which can used to tighten the relaxation by cutting off parts of the relaxation polytope.

\spara{Knapsack constraint.}
We now proceed to describe how to express the
linear knapsack-type constraint $\vweight^\top \bfx \leq \budget$
with respect to the new variable \matrixX.
One straightforward way is to apply the constraint on the diagonal elements of $\matrixX$,
leading to
\begin{equation}
\label{equation:diagonal-constraint}
\trace(\Diag(\vweight)\,\matrixX) \leq \budget.
\end{equation}

In order to further tighten the relaxation
we replace constraint~(\ref{equation:diagonal-constraint})
by a tighter one, which is due to Helmberg et al.~\cite{Helmberg96asemidefinite}.
In their work, they show that the \textit{square representation} constraint
\[
\trace(\vweight\,\vweight^\top\matrixX) \leq \budget^2
\]
is tigher than constraint~(\ref{equation:diagonal-constraint}).

\spara{The resulting \sdp relaxation.}
Putting everything together,
our \sdp relaxation becomes
\begin{equation}
\tag{\sdpqbk}\label{problem:SDP-QBK}
\begin{aligned}
     &\max & & \trace(\matrixP\,\matrixX)\\
     & \text{subject to} & & \trace(\vweight\,\vweight^\top\matrixX) \leq \budget^2 \\
     &&& \matrixX - \bfx\,\bfx^\top \succeq 0 \\
     &&& \diag(\matrixX)=\bfx.
\end{aligned}
\end{equation}

The \sdpqbk problem is convex and can be solved efficiently
by readily available packages.

\spara{Rounding.}
Let $\opt = (\matrixXopt,\bfxopt)$ be an optimal
solution of \sdpqbk, obtained by a convex-optimization solver.

The last step is to round the optimal solution \opt
to a binary vector \bfxbin that is a feasible solution for the \qbk problem.

In order to derive such a binary solution for \qbk,
we follow the randomized-rounding approach
proposed by Luo et al.~\cite{luo2010semidefinite}:
Consider a semi-definite program (\bsdp) over a binary vector \bfx,
and its relaxation (\sdpr) over a lifted variable matrix \matrixX,
where $\matrixX = \bfx\,\bfx^\top$.
Let \matrixXopt be the optimal solution to (\bsdp),
and consider a random  vector $\bfz$ drawn from a Gaussian
distribution with zero mean covariance \matrixXopt,
or, $\bfz \sim \Gaussian(\bfzero, \matrixXopt)$.
It can be shown that \bfz defines a distribution
for which the quadratic objective of (\sdpr) is maximized
and its quadratic constraints are satisfied \emph{in expectation}.
Then, a feasible binary solution \bfxbin to (\bsdp) can be constructed as follows
\begin{enumerate}
\item solve (\sdpr) to find optimal solution \matrixXopt;
\item draw $\bfz \sim \Gaussian(\bfzero, \matrixXopt)$;
\item round \bfz to a binary \bfxbin;
\item repeat (3) until the constraints of (\bsdp) are satisfied.
\end{enumerate}
As shown by Luo et al.~\cite{luo2010semidefinite},
in certain cases,
this randomized rounding technique can give solutions with
a provable quality guarantee.
This is clearly not the case in our problem,
as we have shown that it is inapproximable. 
However, the randomized rounding technique can still be used
as a powerful heuristic in the context of \sdp relaxation.

In our case, after solving the relaxed problem \sdpqbk
and obtaining an optimal solution $\opt = (\matrixXopt,\bfxopt)$,
we draw a random vector $\bfz \sim \Gaussian(\bfxopt, \matrixXopt - \bfxopt\,{\bfxopt}^{\top})$,
The coordinates of \bfxopt are between 0 and 1,
and the coordinates of \bfz are truncated to be between 0 and 1.
The vector \bfz is rounded to a binary vector \bfxbin
using \emph{randomized rounding},
i.e., ${\bfxbin}_i$ is set to 1 with probability equal to ${\bfz}_i$.

The optimal solution $\opt = (\matrixXopt,\bfxopt)$ of \sdpqbk
maximizes the stochastic quadratic objective
\[
\expectation_{\bfz \sim \Gaussian(\bfxopt, \matrixXopt - \bfxopt\,{\bfxopt}^{\top})}
\left[ \bfz^\top\matrixP\,\bfz \right],
\]
which is also the stochastic version of the objective for \qbk.
The optimal solution $\opt = (\matrixXopt,\bfxopt)$ also satisfies the constraint
\[
\expectation_{\bfz \sim \Gaussian(\bfxopt, \matrixXopt - \bfxopt\,{\bfxopt}^{\top})}
\left[ \bfz^\top \vweight\, \vweight^\top \bfz \right] \le \budget^2.
\]
Thus, the binary vector \bfxbin obtained by the randomized rounding
satisfies the constraint
$
\bfxbin^\top \vweight\, \vweight^\top \bfxbin  \le \budget^2
$
in expectation.
In addition, until \bfxbin satisfies the knapsack constraint of the \qbk problem
$
\vweight^\top \bfxbin  \le \budget
$
new randomized binary vectors \bfxbin are drawn.
The resulting algorithm \sdpalgo is shown as Algorithm~\ref{algorithm:sdp}.

\begin{algorithm}[t]
\begin{small}
\SetKwInOut{Input}{input}
\SetKwInOut{Output}{output}
\SetKwRepeat{Do}{do}{while}
\SetKwComment{Comment}{//\ }{}
\caption{\sdpalgo}
\label{algorithm:sdp}
\Input{matrix \matrixP; node weights \vweight; budget \budget; number of iterations \iter}
\Output{indicator vector \bfxbin}
solve \sdpqbk and obtain $\opt \leftarrow (\matrixXopt,\bfxopt)$\;
form covariance matrix $\Sigma \leftarrow\matrixXopt-{\bfxopt}{\bfxopt}^\top$\;
compute the Cholesky factorization $\Sigma=\matrixV\,\matrixV^\top$\;
initialize $\bfxbin\leftarrow\bfzero$ and $f\leftarrow 0$\;
\For{$i\leftarrow1,\ldots,\iter$}{
sample $\bfz \leftarrow \bfxopt + \matrixV\,\bfr$, where $\bfr \sim\Gaussian(\bfzero, \matrixI)$\;
\Comment{$\bfz \sim \Gaussian(\bfxopt, \matrixXopt - \bfxopt\,{\bfxopt}^{\top})$}
 \Do{$\vweight^\top\bfxbin' > \budget$}{
  $\bfxbin'\leftarrow \mathrm{randomized\_rounding}(\bfz)$\;
 }
\If{$f < \bfxbin' \matrixP\,\bfxbin'^\top$}{$\bfxbin\leftarrow\bfxbin'$ and $f\leftarrow\bfxbin' \matrixP\,\bfxbin'^\top$\;}
}
\Return \bfxbin\;
\end{small}
\end{algorithm}

\subsection{Glover's linearization}

An alternative way to handle the difficulty of quadratic programs
and solve them efficiently is to perform \textit{linearization},
i.e., reformulate the quadratic program as a linear program using auxiliary variables and constraints.
A concise way to linearize a 0--1 quadratic programs is Glover's linearization~\cite{Glover75}.
According to this technique we set
$z_i=\xelement_i\sum_{j=1}^\novertices \Pelement_{ij}\xelement_j$ and reformulate our program as:
\small
\begin{align}
&\text{maximize} & & \sum_{i=1}^\novertices z_i \nonumber\tag{\glover}\label{problem:Glover}\\
&\text{subject to} & &  \bfx \in \{0,1\}^\novertices\nonumber\\
&&&  \sum_{i=1}^\novertices \xelement_i\leq \budget  \nonumber\\
&&&   \xelement_i L_i\leq z_i \leq \xelement_i U_i, & i=1,\ldots, \novertices \label{ineq:glover1}\\
&&&   \sum_{j=1}^\novertices\Pelement_{ij} \xelement_j-U_i(1-\xelement_i)\leq z_i,
& i=1,\ldots,\novertices,\label{ineq:glover2}\\
&&&   z_i\leq \sum_{j=1}^\novertices\Pelement_{ij} \xelement_j-L_i(1-\xelement_i),
& i=1,\ldots,\novertices,\label{ineq:glover3}
\end{align}
\normalsize
where $L_i$ and $U_i$ are lower and upper bounds for $z_i$, respectively.
Observe that we can easily obtain such bounds;
for each $L_i$ it suffices to set $\xelement_j=0$ if $\Pelement_{ij}<0$ and
$\xelement_j=1$ if $\Pelement_{ij}\ge 0$,
while to
obtain $U_i$ it suffices to set $\xelement_j=0$ if if $\Pelement_{ij}\ge 0$, and
$\xelement_j=1$ if $\Pelement_{ij}<0$.

Inequalities (\ref{ineq:glover1})--(\ref{ineq:glover3})
enforce the following equivalence between problems \qbk and \glover.
If $\xelement_i =0$ for some $i$, then (\ref{ineq:glover1}) ensures that $z_i=0$
and (\ref{ineq:glover2})+(\ref{ineq:glover3}) are redundant.
If $\xelement_i=1$ for some $i$, then
(\ref{ineq:glover2})+(\ref{ineq:glover3})
ensure that $z_i = \sum_{j=1}^\novertices \Pelement_{ij}\xelement_j$
and (\ref{ineq:glover1}) is redundant.
In either case, $z_i = \xelement_i\sum_{j=1}^\novertices \Pelement_{ij}\xelement_j$ for each $i$.


After formulating \glover,
we solve the continious relaxation of this integer program,
and we obtain a bound on the initial program. The obtained fractional solution $\bfx$ is converted into a binary vector $\bfxbin$ using a similar procedure as before: we repeatedly perform randomized rounding, where each $\xbinelement_i$ is set to 1 with probability equal to $\xelement_i$, until we obtain a feasible solution.

\subsection{Greedy algorithms}

Solving an \sdp problem, up to a desirable accuracy $\epsilon$,
requires time polynomial in the problem size $n$ and $\log\frac{1}{\epsilon}$.
However, \sdp solvers are using expensive interior-point methods
with running time $\tilde{\bigO}(n^3)$.
Thus, the \sdpalgo algorithm discussed in the previous section
is expected to produce solutions of high quality,
but it is not scalable to problem instances of large size.

In this section, we present two greedy algorithms for the \qbk problem,
which scale linearly to the size of the input graph.
Additionally, as we will see in our experimental evaluation,
the greedy algorithms yield solutions of extremely high quality, in practice.

\smallskip
\noindent
\textbf{Simple greedy.}
The first scalable algorithm is a simple greedy (\heuristicalgo),
which takes advantage of the structural properties of \matrixP.
Recall from Equation~(\ref{equation:P}) that the entries of \matrixP
incorporate information about the structure of the social network,
as well as the exposure values of the neighbors of each node.

Notice that
the diagonal entries $\Pelement_{ii}$ of matrix \matrixP
take values in $[-\Delement_{ii},\Delement_{ii}]$,
where $\Delement_{ii}$ is the weighted degree of node~$i$,
while the rest of the matrix is sparse.
Also observe that in the 0--1 quadratic function
$\bfx^\top\matrixP\,\bfx$, setting an element of \bfx to $1$
means that the corresponding diagonal element of $\matrixP$ is selected.
Therefore, it is beneficial to select first those indices that
correspond to the highest diagonal values of $\matrixP$.
In order to account for the node weight \vweight
we select the node with the highest ratio $\Pelement_{ii}/\vwelement_i$,
that is, the most cost-effective node for its contribution to the objective.
%

Altogether, the \heuristicalgo algorithm selects nodes
in descending order of value $\Pelement_{ii}/\vwelement_i$,
while the total weight of selected nodes
($\sum_i \vwelement_i \xelement_i$) does not exceed the budget~\budget.

\smallskip
\noindent
\textbf{Iterative greedy.}
An obvious drawback of \heuristicalgo
is that nodes are selected independently,
and thus, selected nodes may have the effect of canceling each other.

Our second greedy strategy (named iterative greedy, or simply \greedyalgo)
overcomes this drawback
by selecting nodes iteratively and evaluating the gain in the objective function
for each new node.
The \greedyalgo algorithm first generates a feasible solution \bfx
by initially setting all nodes to 0.
Then it iteratively sets the value of a variable from 0 to 1,
so as to achieve the highest gain in the objective value,
normalized for node cost, i.e.,
it selects the node $i$ that achieves the highest value of the ratio
$\frac{\bfx'^\top\matrixP\,\bfx'}{\vwelement_i}$,
where $\bfx'$ differs from the current solution \bfx
as to having its $i$-th coordinate equal to 1 instead of 0.
The algorithm continues adding nodes
while the total weight of selected nodes
($\sum_i \vwelement_i \xelement_i$) does not exceed the budget~\budget.

To further explore the search space,
and allow the possibility of recovering from a bad choice during the greedy selection,
we enhance the algorithm with an additional local-search step.
According to this,
a node in the current solution is selected at random,
removed from the solution,
and
other nodes are selected greedily to replace the removed node.
The local-search step is repeated for a given number of iterations \iter.
The \greedyalgo algorithm returns the best solution found during its execution.

The \greedyalgo algorithm is described in Algorithm~\ref{algorithm:greedy}.

\begin{algorithm}[t]
\begin{small}
\SetKwInOut{Input}{input}
\SetKwInOut{Output}{output}
\SetKwRepeat{Do}{do}{while}
\SetKwComment{Comment}{//\ }{}
\caption{\greedyalgo}
\label{algorithm:greedy}
\Input{matrix \matrixP; node weights \vweight; budget \budget; iterations \iter}
\Output{indicator vector \bfxbest}
initialize $\bfxbest\leftarrow\bfzero$, $\bfx\leftarrow\bfzero$ and $f\leftarrow0$\;
\For{$i \leftarrow 1,\ldots,\iter$}{
  $\bfx'\leftarrow\bfx$\;
  \While{$\vweight^\top\bfx'\le\budget$}{
    $\bfx \leftarrow \bfx'$\;
    $j^* \leftarrow \arg\max_j \left\{ \frac{\bfx'^\top\matrixP\,\bfx'}{\vwelement_j}
            \mid \bfx'\leftarrow\bfx \mbox{ and } \xelement'_{j}\leftarrow 1 \right\}$ \label{line:argmax}\;
    $\xelement'_{j^*}\leftarrow1$\;
  }
  \If{$f < \bfx^\top \matrixP\,\bfx$}{$\bfxbest\leftarrow\bfx$ and $f\leftarrow\bfx^\top \matrixP\,\bfx$\;}
  $r \leftarrow \mathit{random}\,\{ j \mid \xelement_j = 1 \}$\;
  $\xelement_r \leftarrow 0$\;
}
\Return \bfxbest\;
\end{small}
\end{algorithm}

To analyze the running time of \greedyalgo,
consider the computation of the value $\bfx'^\top\matrixP\,\bfx'$.
A na\"ive implementation uses vector-matrix multiplication
and results in complexity $\bigO(\novertices^2)$.
However, we can improve the running time considerably,
by observing that multiplying a matrix with a binary vector
is equivalent to selecting its rows or columns that correspond to indices with value 1 in the vector.
Therefore we can compute the updated
value $\bfx'^\top\matrixP\,\bfx'$
by selecting a single column and row for the new index,
and summing
the nodes that are indexed by the current index set.
Assuming that a solution has at most $\ell$ nodes, the cost is $\bigO(\ell)$.
The total cost in selecting the best index
is $\bigO(\novertices\ell)$.
Overall, the total running time of the \greedyalgo is $\bigO(\novertices\ell^2\iter)$.
In typical scenarios we can assume $\ell,\iter<<\novertices$,
making the algorithm very efficient.

\subsection{Mixed integer quadratic programming}

For problem instances of small size,
we can solve \qbk optimally,
using a mixed-integer quadratic programming package.
Although the computational complexity of such a method
is exponential in the worst case,
powerful general-purpose solvers may work well for real-world (not worst case) inputs.

By solving the problem optimally on small-size datasets,
we can evaluate our more scalable techniques
by checking how far off they are from the optimal solution.
In our experiments we use {\sc\large cplex},
a standard  mixed-integer quadratic programming solver.

\subsection{Upper bounds}
\label{section:upper-bounds}

In this section we derive three upper bounds
for problem \qbk.
Our bounds are applicable to the special case of
all nodes having the same cost ($\vweight=\bfone$).
This is 
equivalent to setting a cardinality constraint on vector \bfx.
The bounds also hold in the case that $\vwelement_i\geq 1$,
but they may not be as tight in that case.
The three bounds we present differ in computational complexity and tightness.

Computing a tight upper bound for our problem has several benefits.
First we can compare the upper bound with the value of the solution obtained by a method
and having an estimation of the approximation for a particular problem instance.
Second, a bound can be used to speed up some of our algorithms,
e.g.,
in a branch-and-bound routine when computing the optimal solution,
or as a cutting plane in the \sdp relaxation algorithm.

The Rayleigh theorem for Hermitian matrices $\matrixM$,
provides an upper bound for the quantity $\bfx^\top \matrixM\, \bfx$
based on the maximum eigenvalue $\lambdamax(\matrixM)$ of \matrixM:
\[
\bfx^\top \matrixM\,\bfx \leq \lambdamax(\matrixM)\,\bfx^\top\bfx.
\]
The constraint $\bfone^\top \bfx \leq \budget$ implies that $\bfx^\top\bfx \leq \budget$,
and therefore:
\[
\bfx^\top\matrixP\,\bfx \leq \lambdamax(\matrixP)\,\bfx^\top\bfx \leq \lambdamax(\matrixP)\,\budget.
\]

We can estimate the dominant eigenvalue $\lambdamax(\matrixP)$
iteratively.
Taking advantage of the sparsity of $\matrixP$,
the complexity of estimating this bound is $\bigO(\sparsity\iter)$,
where \sparsity is the number of non-zero elements in \matrixP
and \iter is the number of iterations required for convergence.

\smallskip
Our second bound can be computed in time $\bigO(\novertices)$,
based on
the following lemma, which is a consequence of
Gerschgorin's~theorem.

\begin{lemma}
\label{lemma:Gerschgorin}
$\lambdamax(\matrixM) \leq
\gbound = \max_i\left\{ \Melement_{ii} + \sum_{i\neq j}|\Melement_{ij}| \right\}.$
\end{lemma}

Recall the definition of \matrixP,
and observe that
$\Pelement_{ii}=-\bfs^\top\laplacian_{\cdot i}\,\selement_i$ and
$\Pelement_{ij}= \selement_{i}\Lelement_{ij}$, for $i\neq j$,
where $\laplacian_{\cdot i}$ is the $i$-th column of the Laplacian matrix,
and $\Lelement_{ij}$ the $(i,j)$ entry.
It follows that
$\Pelement_{ii} + \sum_{i \neq j} |\Pelement_{ij}| = \Pelement_{ii} + \Delement_{ii}$,
where $\Delement_{ii}$ is the weighted degree of node $i$.
From Lemma~\ref{lemma:Gerschgorin} it follows that
$\lambdamax(\matrixP)\leq \gbound =\max_i\{\Pelement_{ii}+\Delement_{ii}\}$ and consequently
\begin{equation*}
\label{eq:eigenconstraint}
\bfx^\top\matrixP\,\bfx \leq \budget \gbound.
\end{equation*}


\smallskip
The third bound is based on the following observation:
given upper bounds on the rows of \matrixP,
due to the cardinality constraint on \bfx,
the value of the objective function can be at most the sum of the \budget highest row-upper bounds.
Accordingly, an upper bound on each row can be obtained by summing the top \budget nonnegative nodes.
This bound is more expensive than the other two,
as it requires $\bigO(\novertices^2)$ operations, but we expect it to be tighter.

\section{Extension to the continuous case}
\label{section:continuous}

Here we extend our problem formulation to the continuous case:
we want to select \budget
individuals and modify their exposure to some value in the interval $[-1,1]$.
The goal is again to maximize the diversity index.
The choice of the algorithm is which individuals to select
and a recommended exposure level for each one of them.

\begin{problem}\label{problem:problem2}
Given a graph $G=(V,E)$,
an exposure vector \bfs,
and an integer \budget,
identify a sparse vector \bfx with \budget
non-zero nodes $\xelement_i \in [-1+\selement_i,1+\selement_i]$,
such that the resulting \textit{diversity index} $\eta(\graph,\bfx-\bfs)$ is maximized.
\end{problem}
By setting $\bfy=\bfx-\bfs$ we obtain the following problem:
\begin{equation}
\begin{aligned}
 \label{eq:UP}
     &\max & & \bfy^\top \laplacian\,\bfy, \\
     & \text{subject to} & & \|\bfy\|_{\infty}\leq 1, \\
     &&& \card(\bfy+\bfs) \leq \budget.
\end{aligned}
\end{equation}
This problem is difficult to solve due to the non-convex cardinality constraint.
Again, we will use  a semidefinite relaxation to turn the non-convex problem into a convex one.
First we 
introduce the variable $\matrixY=\bfy\,\bfy^\top$.
We can represent the constraint $\|\bfy\|_{\infty}\leq 1$ either with the linear constraint   $\diag(\matrixY) \leq \bfone$ or with the tighter but quadratic constraint $\|\matrixY\|_{\infty}\leq1$, where $\|\matrixM\|_{\infty}=\max_{i,j=1,...,n}|M_{ij}|$.

Next we describe how to rewrite the non-convex cardinality constraint.
We observe that:
\[
\|\bfy+\bfs\|_1^2= \bfone^\top|\matrixY+\bfs\,\bfs^\top+2\bfy\,\bfs^\top|\,\bfone.
\]

Additionally, 
for every vector~\bfu with $\card(\bfu)=k$
it is
$\|\bfu\|^2_1 \leq k^2 \|\bfu\|^2_\infty$.
Thus, we can replace
$\card(\bfy+\bfs) \leq \budget^2$
with
$\bfone^\top|\matrixY+\bfs\,\bfs^\top+2\bfy\,\bfs^\top|\,\bfone\leq \budget^2\|\bfy+\bfs\|^2_\infty \leq 4\budget^2$.
The last inequality follows since
$\|\bfy\|_{\infty}\leq 1$ and $\|\bfs\|_{\infty}\leq 1$ implies
$\|\bfy+\bfs\|_{\infty}\leq2$.
Here, $|\matrixA|$ denotes the element-wise absolute value of matrix \matrixA.

Finally, we relax the non-convex $\matrixY=\bfy\,\bfy^\top$ constraint by a matrix inequality,
which relies on the Schur complement, as we have shown before.
The resulting relaxation is:

\begin{equation}
\begin{aligned}
 \label{eq:RP}
     &\max & & \trace(\laplacian\, \matrixY) \\
     & \text{subject to} & & \|\matrixY\|_{\infty}\leq1\\
     &&& \bfone^\top|\matrixY+\bfs\,\bfs^\top+2\,\bfy\,\bfs^\top|\,\bfone \leq 4\budget^2 \\
     &&& \matrixY-\bfy\,\bfy^\top \succeq 0.
\end{aligned}
 \end{equation}

The optimal value of the quadratic program (\ref{eq:RP})
is an upper bound on the optimal value of program (\ref{eq:AP}).
The optimal solution \matrixY will not always be a rank-one matrix
but we can truncate it and keep only its dominant eigenvector.


The semidefinite program (\ref{eq:RP}) can now be solved efficiently using off-the-shelf methods. 

\section{Experiments}
\label{section:experiments}

In this section, we present an experimental evaluation of the algorithms we presented.
The goal of our experiments is threefold:
First, we want to compare the performance of the algorithms in terms of the achieved value of the objective function.
Second, we want to evaluate the scalability of the algorithms.
Finally, we want to investigate the factors affecting the performance of the algorithms.

All experiments are conducted on an HPC machine with 8-cores and 32\,GB of RAM.

\smallskip\noindent{\bf Datasets.}
We consider five datasets representing different types of social networks.
We use networks where each node is associated with a value between $-1$ and $1$,
which we assume that reflects its exposure.
We consider the following datasets:

\smallskip\noindent
{{\karate}}:\footnote{https://networkdata.ics.uci.edu/data.php?id=105}
The well-known dataset representing a social network
of a karate club at a US university in the 1970s.
The social network is partitioned into two distinct equal-sized communities.

\smallskip\noindent
{{\books}}:\footnote{https://networkdata.ics.uci.edu/data.php?id=8}
A network of books about US politics,
sold by 
{\tt amazon.com}.
Edges 
represent frequently co-purchased books.
Books are classified as \textit{Liberal} (43), \textit{Conservative} (49), and \textit{Neutral} (13).
Neutral books are randomly assigned to one of the two communities.

\smallskip\noindent
{{\blogs}}:\footnote{https://networkdata.ics.uci.edu/data.php?id=102}
A directed network of hyperlinks between weblogs on US politics recorded in
2005~\cite{Adamic:2005:PBU:1134271.1134277}.
Blogs are classified as either \textit{Liberal} or \textit{Conservative}.
We disregard edge directions and keep the largest connected component.
The resulting dataset contains two communities with 636 and 586 nodes each.

\smallskip\noindent
{{\elections}}:
A network of twitter followers of D.\ Trump and H.\ Clinton
collected in the end of 2016.
We consider two communities of users, partitioned by the usage of hashtags
{\tt \#maga} 
and {\#imwithher}. 
We keep the largest connected component and iteratively prune nodes
to guarantee that every node has degree greater than 1.

\smallskip\noindent
{{\twitterbig}}:
A large network of twitter users collected between 2011 and 2016,
filtered for keywords related to three controversial topics:
\emph{gun control}, \emph{abortion}, and \emph{obamacare}~\cite{Lahoti17}.
For the exposure of the users we use the ideology scores estimated by
Barber\'a et al.~\cite{barbera2014birds}. We only expect the two greedy
algorithms to scale on this dataset, therefore in order to evaluate all our algorithms 
we also generate a smaller dataset. We rank the nodes of the network according to their pagerank values, and
keep the largest connected component formed from the subgraph of the top-100 nodes.
We refer to this smaller dataset as {\twitter}. 

\smallskip
To evaluate our algorithms in networks that are already diverse and
there is no latent community structure,
we create a version of networks {{\karate}}, {{\books}}, and {\twitter},
where nodes are assigned a random exposure value.
The resulting networks are called {\karated}, {\booksd}, and {\twitterd},
respectively.

Table \ref{dataset-stats} shows the statistics of our datasets.
All networks are treated as undirected.
All edge weights and node costs are set to~1.

\begin{table}
	\centering
	\caption{\label{dataset-stats}Dataset Statistics}
	\vspace{-3mm}
	\resizebox{\columnwidth}{!}{%
	\begin{tabular}{@{}lrrrrrr@{}}
		\toprule
		Dataset & Nodes & Edges & Avg Degree & Positive & Negative & $\eta$\\
		\midrule
		\karate\ & 34 & 78 & 4.58 & 17 & 17 & 10 \\
		\karated\ & 34 & 78 & 4.58 & 18 & 16 & 43 \\
		\books\ & 105 & 441 & 8.40 & 43 & 49 & 35\\
		\booksd\ & 105 & 441 & 8.40 & 54 & 51 & 224\\
		\twitter\ & 80 & 1\,403 & 17.53 & 25 & 55 & 90 \\
		\twitterd\ & 80 & 1\,403 & 17.53 & 42 & 38 &  710\\
		\blogs\ & 1\,222 & 16\,717 & 27.36 & 636 & 586 & 1\,419\\
		\elections \ & 18\,893 & 269\,696 &  14.27 & 6\,612 & 12\,281 & 28\,164\\
		\twitterbig \ & 200\,073 & 4\,009\,548 & 50\,038.04 & 81\,793 & 118\,280 & 251\,450\\
		\bottomrule
	\end{tabular}
	}
\vspace{-1mm}
\end{table}

\smallskip\noindent{\bf Performance evaluation.}
We first evaluate the algorithms with respect to the diversity-index score they achieve.
{\sdpalgo} is the \sdp-based algorithm,
{\gloveralgo} is the linearization algorithm,
{\greedyalgo} and {\heuristicalgo} are the two greeedy algorithms,
while
{\exactalgo} is the exact algorithm.

Table~\ref{objective-value} shows the results obtained by the algorithms on all datasets.
For the smaller datasets, {\karate}, {\books} and {\twitter},
we set $\budget=0.1\,\novertices, 0.2\,\novertices, \novertices$,
while for the larger datasets we set $\budget=0.1\,\novertices$ for {\blogs},
and $\budget=0.01\,\novertices$ for {\elections}. For the largest dataset {\twitterbig}, we set $\budget=0.001\,\novertices$. 

\begin{table}[t]
	\centering
	\caption{Solution quality and bounds from the relaxations}
	\vspace{-3mm}
	\label{objective-value}
	\resizebox{\columnwidth}{!}{%
	\begin{tabular}{@{}lrrrrrr@{}}
		\toprule
		Dataset & \budget & \exactalgo & \sdpalgo & \gloveralgo & \greedyalgo & \heuristicalgo \\
		\midrule
		{\karate}\ & {0.1\,\novertices}\ & 46 & 46 (46.43) & 46 (52.28) & 46 & 46 \\
		& {0.2\,\novertices} & 56 & 56 (59.13) & 54 (69.05) & 56 & 51 \\
		& {\novertices} & 61 & 61 (63.48) & 52 (78.00) & 57 & 51 \\
		{\karated}\ & {0.1\,\novertices} & 50 & 50 (53.69) & 49 (65.85) & 50 & 50 \\
		& {0.2\,\novertices} & 55 & 55 (60.74) & 50 (79.84) & 53 & 52 \\
		& {\novertices} & 61 & 61 (63.89) & 50 (93.00) & 55 & 48 \\
		{\books}\ & {0.1\,\novertices} & 207 & 207 (207.81) & 207 (235.90) & 207 & 207 \\
		& {0.2\,\novertices} & 264 & 262 (272.26) & 249 (330.01) & 264 & 248 \\
		& {\novertices} & 309& 306 (318.43) & 267 (447.00) & 298 & 253 \\
		{\booksd}\ & {0.1\,\novertices} & 265 & 262 (272.95) & 249 (328.32) & 263 & 252 \\
		& {0.2\,\novertices} & 285 & 281 (298.48) & 280 (388.82) & 284 & 254 \\
		& {\novertices} & 309 & 307 (318.44) & 282 (497.50) & 286 & 243 \\
		{\twitter}\ & {0.1\,\novertices}  & 425 & 425 (425.19) & 424 (479.52) & 425 & 424 \\
		& {0.2\,\novertices} & 599 & 599 (601.38) & 589 (791.55) & 599 & 592 \\
		& {\novertices}  & $-$ & 793 (804.21) & 733 (1\,406.00) & 790 & 647 \\
		{\twitterd}\ & {0.1\,\novertices} & 743 & 742 (752.94) & 722 (917.99) & 742 & 715 \\
		& {0.2\,\novertices} & $-$ & 757 (775.53) & 729 (1\,071.32) & 761 & 715 \\
		& {\novertices}  & $-$ & 793 (804.21) & 775 (1\,496.00) & 766 & 737 \\
		{\blogs}\ & {0.1\,\novertices} & $-$ & $-$ & 9\,878 (12\,659.06) & 9\,879 & 8\,889 \\
		{\elections}\ & $0.01n$ & $-$ & $-$ & $-$  & 117\,950 & 117\,482  \\
		{\twitterbig}\ & $0.001n$ & $-$ & $-$ & $-$  & $-$ & 1\,678\,753  \\
		\bottomrule
	\end{tabular}
	}
\vspace{-1mm}
\end{table}

We observe that {\sdpalgo} is the best-performing algorithm:
it finds solutions of quality very close to that of {\exactalgo}, which is optimal.
Especially surprising is the performance of {\greedyalgo},
which is almost equal to {\sdpalgo}.
It even outperforms {\sdpalgo} slightly in some instances.
On the other hand, {\greedyalgo} performs less well for $\budget=n$,
which is expected, given its greedy nature.

It is important to note that {\exactalgo} terminates in reasonable time
only for networks of up to $100$ nodes.
We also observe that the \sdp relaxation is tight and achieves upper bounds very close to the optimal value
(always less than 1.007 times the optimal).
{\gloveralgo}, on the other hand,
does not give tight relaxations: its upper bounds can get quite off.
Finally, {\heuristicalgo} manages to achieve good performance for small-size instances,
due to picking first the high diagonal elements,
but it fails to give good solutions for larger instances.

\begin{table}[t]
\centering
\caption{Upper bounds}
\vspace{-3mm}
\label{Bounds}
{\footnotesize
\begin{tabular}{@{}lrrrrr@{}}
\toprule
Dataset & k & Optimal & Bound1 & Bound2 & Bound3  \\
\midrule
{\karate}\ & {0.1\,\novertices}  & 46 & 122 & 63.62 & 57 \\
& {0.2\,\novertices} & 56 & 262 & 130.63 & 83 \\
        & {\novertices} & 61 & 934 & 452.32 & 145  \\
{\karated}\ & {0.1\,\novertices}  & 50 & 99 & 76.43 & 59 \\
& {0.2\,\novertices} & 55 & 169 & 118.23 & 90 \\
        & {\novertices} & 61 & 570 & 302.43 & 145 \\
{\books}\ & {0.1\,\novertices} & 207 & 535 & 297.6 & 245 \\
 & {0.2\,\novertices} & 264 & 1\,035 & 560.20 & 387 \\
       & {\novertices} & 309 & 5\,235 &  2\,766.04 & 857  \\
{\booksd}\ & {0.1\,\novertices} & 265 & 552 & 363.26 & 338 \\
&{0.2\,\novertices} & 285 & 872 & 494.53 & 439 \\
       & {\novertices} & 309 & 3\,528 &  1\,249.18 & 778 \\
{\twitter}\ & {0.1\,\novertices} & 425 & 890 & 498.27 & 481 \\
& {0.2\,\novertices} & 599 & 1\,590 & 855.50 & 807 \\
{\twitterd}\ & {0.1\,\novertices} & 743 & 1\,176 & 1\,035.32 & 809 \\
\bottomrule
\end{tabular}
}
\vspace{-1mm}
\end{table}

In addition, we evaluate the quality of the three upper bounds
(Section~\ref{section:upper-bounds}).
Table~\ref{Bounds} shows the results.
Bound3 is the most expensive to compute, but is also tightest.
%
On the other hand,
Bound1 is the cheapest to compute, by it can get quite bad.
We also observe that is more tight for diverse networks.
This is due to the impact of the diagonal elements of \matrixP on the computation of the bound:
the diagonal elements are smaller for diverse networks.
Finally, Bound2 is fairly tight for $\budget=0.1\,\novertices$ 
but it gets worse for $\budget=\novertices$.
In general, we observe that for all bounds the value is much closer to the optimal for small instances,
which is somewhat expected.
It is worth noting that for the case $\budget=\novertices$,
despite the fact that the optimal diversity-index value is the same no matter the initial assignment of exposures
(since all exposures can be changed),
the bounds obtained are different.


\begin{table}[t]
	\centering
	\caption{Running times (in seconds)}
	\vspace{-3mm}
	\label{running-time}
	\resizebox{\columnwidth}{!}{%
	\begin{tabular}{@{}lrrrrrr@{}}
		\toprule
		Dataset & \budget & \exactalgo & \sdpalgo & \gloveralgo & \greedyalgo & \heuristicalgo \\
		\midrule
		{\karate}\ & {0.1\,\novertices} & 0.093 & 1.355 & 0.814 & 0.009 & 0.002 \\
		& {0.2\,\novertices} & 0.274 & 1.326 & 0.575 & 0.018 & 0.001 \\
		& {\novertices} & 1.620 &  1.820 & 0.587 & 0.035 & 0.001 \\
		{\karated}\ & {0.1\,\novertices} & 0.172 & 1.436 & 0.692 & 0.010 & 0.002 \\
		& {0.2\,\novertices} & 0.275 & 1.271 & 0.613 & 0.019 & 0.002 \\
		& {\novertices} & 0.329 & 1.460 & 0.558 & 0.036 & 0.001 \\
		{\books}\ & {0.1\,\novertices} & 0.098 & 158.297 & 6.259 & 0.078 & 0.002 \\
		&{0.2\,\novertices} & 0.334 & 165.299 & 5.157 & 0.154 & 0.002 \\
		& {\novertices} & 2.543 & 213.720 &  4.744 & 0.266 & 0.006 \\
		{\booksd}\ & {0.1\,\novertices} & 0.503 & 146.344 & 6.138 & 0.123 & 0.002 \\
		& {0.2\,\novertices} & 1.493 & 138.263 & 6.225 & 0.150 & 0.002 \\
		& {\novertices} & 3.726 & 188.170 &  4.744 & 0.253 & 0.006 \\
		{\twitter}\ & {0.1\,\novertices} & 0.855 & 44.813 & 3.745 & 0.042 & 0.001 \\
		& {0.2\,\novertices} & 71.362 & 50.523 & 3.139 & 0.083 & 0.002 \\
		\ &{\novertices} & $>$7200 & 56.670 & 2.870 & 0.086 & 0.002 \\
		{\twitterd}\ & {0.1\,\novertices} & 40.284 & 40.972 & 3.687 & 0.041 & 0.002 \\
		& {0.2\,\novertices} & $>$7200 & 39.720 & 2.980 & 0.077 & 0.001 \\
		\ & {\novertices} & $>$7200 & 42.811 & 3.007 & 0.072 & 0.002 \\
		{\blogs}\ & {0.1\,\novertices} & $-$   &  $-$      & 947.980 & 10.070 & 0.103 \\
		{\elections}\ & {0.01\,\novertices} & $-$ & $-$ & $-$  & 333.727 & 12.961  \\
		{\twitterbig}\ & {0.001\,\novertices} & $-$ & $-$ & $-$  & $-$ & 3000.676  \\
		\bottomrule
	\end{tabular}
	}
\vspace{-1mm}	
\end{table}

\smallskip\noindent{\bf Scalability.}
We also perform a scalability analysis of the algorithms,
shown in Table~\ref{running-time}.
We are able to run all algorithms for the smaller datasets,
{\karate}, {\books}, and {\twitter},
although {\exactalgo} did not terminate after two hours on
{\twitter} for $\budget=\novertices$ and \twitterd for $\budget=0.2\,\novertices$.
For the {\blogs} dataset, {\gloveralgo}, {\greedyalgo}, and {\heuristicalgo} are scalable,
while {\exactalgo} and {\sdpalgo} run out of memory.
{\greedyalgo} and {\heuristicalgo} are very scalable,
and run fast even on big datasets. {\heuristicalgo} scales well even on the very large network, {\twitterbig}. 

All in all,
for the polynomial algorithms the running time is in-line with their theoretical complexity,
while {\exactalgo} is very fast for some instances but
does not terminate within two hours for some other instances.

\smallskip\noindent{\bf Continuous case.}
We also evaluate our \sdp-relaxation for the continuous case
(Section~\ref{section:continuous}).
Due to the presence of two quadratic constraints,
the relaxation terminates only on smallest dataset, \karate.
The running time is 209.95 seconds and the optimal value of the relaxation is 61.17.
However after rounding to a feasible solution to the initial problem,
the values of the solutions drops considerably, to 32.35, lower than that of the discrete problem.
We conclude that while the continuous problem is theoretically interesting,
the proposed algorithm does not perform well---thus, we
pose this problem as an interesting direction for future work.

\smallskip\noindent{\bf Case study.}
We conclude the experiments by taking a closer look at the first five
nodes selected by {\exactalgo} in the {\twitter} dataset.
We characterize the nodes by ranking them according to three measures:
the size of their ``echo chamber,''
defined as the number of their neighbors with the same exposure, 
their centrality, measured by PageRank score, and their degree.
The results are shown in Table~\ref{First_eight}.
We observe that the selected nodes are amongst the highest ranked nodes in all three categories.
It appears that the most important feature when changing the exposure
of an individual is the size of their echo chamber,
This is in line with the observed performance of {\heuristicalgo}
that implements this logic and performs well for small~$\budget$.

\begin{table}[t]
\centering
\caption{Characteristics of the first five nodes selected by \exactalgo on the {\twitter} dataset}
\vspace{-3mm}
\label{First_eight}
{\footnotesize
\begin{tabular}{rrrr}
\toprule
 \# &  \multicolumn{1}{c}{Echo}  & Degree & PageRank  \\
    &  chamber &  &   \\
\midrule
1 &  3 & 6 & 7   \\
2 &  7 & 11 & 11  \\
3 &  8 & 12 & 9  \\
4 &  15 & 13 & 13   \\
5 &  1 & 3 & 3   \\
\bottomrule
\end{tabular}
}
\vspace{-1mm}
\end{table}

\section{Conclusion} 
\label{section:conclusions}

In this paper we considered the problem of diversifying 
user exposure to content in social networks. 
We formally defined the \emph{diversity index} of a social network, 
and formulated the problem of maximizing diversity. 
We showed that the diversity-maximization problem is \np-hard to approximate.
Despite this result, we studied algorithms that in practice offer solutions of high quality, 
including an \sdp-based algorithm, 
an algorithm based on linearization, and two scalable greedy methods. 
Furthermore, we provided several upper bounds with varying tightness-vs.-efficiency trade-off. 
Our experiments with real data demonstrate the effectiveness of our algorithms in the diversity-maximization problem. 
We also introduced a continuous version of our problem, and an \sdp-relaxation. 
Although the continuous version is a relaxation of the discrete problem we studied, 
the proposed \sdp algorithm is not satisfactory neither in terms of quality nor efficiency. 
We consider this variant an interesting and challenging problem to study in the future.  

\spara{Acknowledgments.}
This work has been supported by
three Academy of Finland projects  (286211, 313927, and 317085),
and the EC H2020 RIA project ``SoBigData'' (654024)."

\bibliographystyle{abbrv}
\bibliography{sample-bibliography-brief}

\end{document}